\documentclass[a4paper,fleqn,usenatbib]{mnras}

\usepackage{newtxtext,newtxmath}

\usepackage[T1]{fontenc}
\usepackage{ae,aecompl}

\usepackage{graphicx}	
\usepackage{amsmath}	
\usepackage{amssymb}	
\usepackage{ulem}       
\usepackage{color,xcolor} 

\title[Periodic modulation in PSR J1825$-$0935]{Periodic Q-mode modulation in PSR J1825$-$0935 (PSR B1822$-$09)}

\author[W. M. Yan et al.]{W. M. Yan,$^{1,2}$\thanks{E-mail: yanwm@xao.ac.cn (WMY)}
R. N. Manchester,$^{3}$
N. Wang,$^{1,2,4}$
J. P. Yuan$^{1,2,4}$
\newauthor Z. G. Wen$^{1,2}$
and
K. J. Lee$^{5,6}$
\\
$^{1}$Xinjiang Astronomical Observatory, CAS, 150 Science 1-Street, Urumqi,
Xinjiang, 830011, China\\
$^{2}$Key Laboratory of Radio Astronomy, Chinese Academy of Sciences,
Nanjing 210008, China\\
$^{3}$CSIRO Astronomy and Space Science, Australia Telescope National
Facility, PO Box 76, Epping, NSW 1710, Australia\\
$^{4}$Xinjiang Key Laboratory of Radio Astrophysics, 150 Science 1-Street, Urumqi,
Xinjiang, 830011, China\\
$^{5}$Kavli Institute for Astronomy and Astrophysics, Peking University,
Beijing 100871, China\\
$^{6}$National Astronomical Observatories, Chinese Academy of Sciences,
Beijing 100012, China
}

\date{Accepted XXX. Received YYY; in original form ZZZ}

\pubyear{2019}

\begin{document}
\label{firstpage}
\pagerange{\pageref{firstpage}--\pageref{lastpage}}
\maketitle

\begin{abstract}
PSR J1825$-$0935 (PSR B1822$-$09) switches between radio-quiet
(Q-mode) and radio-bright (B-mode) modes. The Q-mode is known to have
a periodic fluctuation that modulates both the interpulse and the main
pulse with the same period. Earlier investigators argued that the
periodic Q-mode modulation is associated with drifting subpulses.  We
report on single-pulse observations of PSR J1825$-$0935 that were made
using the Parkes 64-m radio telescope with a central frequency of 1369
MHz.  The high-sensitivity observations revealed that the periodic
Q-mode modulation is in fact a periodic longitude-stationary intensity
modulation occurring in the interpulse and the main pulse. The
fluctuation spectral analysis showed that the modulation period is
about $43 P_1$, where $P_1$ is the rotation period of the
pulsar. Furthermore, we confirm that the modulation patterns in the
interpulse and the main pulse are phase-locked. Specifically, the
intensities of the interpulse and the immediately following main pulse
are more highly correlated than for the main pulse and interpulse at any
other lag. Polarization properties of the strong and weak Q-mode states
are different, even for the trailing part of the main pulse which does
not show the periodic intensity modulation. 

\end{abstract}




\section{Introduction} \label{sec:intro}

\begin{figure}
\centering
\includegraphics[angle=0,width=0.45\textwidth]{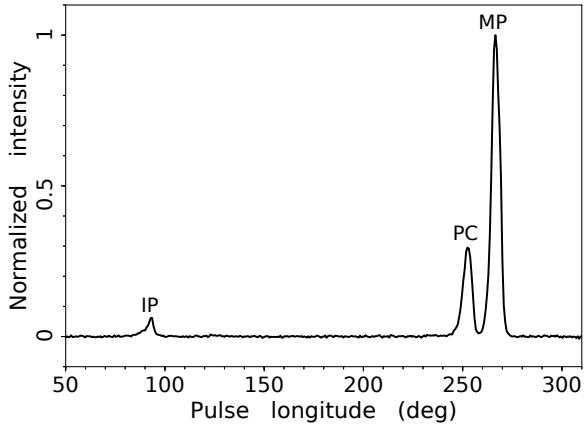}
\caption{The integrated pulse profile of PSR J1825$-$0935 showing the 
         three main components: interpulse (IP), precursor (PC) and
         main pulse (MP). 
         \label{fig:prf_sketch}}
\end{figure}

PSR J1825$-$0935 (PSR B1822$-$09) is a well-known young pulsar which
shows interpulse emission.  As is shown in Figure~\ref{fig:prf_sketch},
the mean pulse profile of PSR J1825$-$0935
consists of three main components: a strong and sharp main pulse (MP),
a precursor (PC) preceding the MP by about $14\fdg5$ of pulse phase,
and a relatively weak interpulse (IP) leading the PC by about
$159\degr$. This pulsar has attracted the attention of
many investigators over the years because it exhibits three
interesting features simultaneously: mode changing, IP-PC
anticorrelation and periodic subpulse modulation.

PSR J1825$-$0935 is known to switch between two stable emission states at radio
frequencies \citep{fwm81,mgb81,gjk+94}. The PC drops to extremely low
intensity in the Q-mode, while in the B-mode, it becomes relatively strong.
\citet{fw82} reported that the mode changing occurs approximately every 5 min. With
a single 8-h GMRT observation, \citet{lmr12} found that the average time between mode
changes is about 7.6 min. From longer multifrequency observations made by three
different radio telescopes (WSRT, GMRT and Lovell), \citet{hkh+17} found an
average of one mode change every 3.5 min, but they did not find evidence for simultaneous
X-ray and radio mode changing. PSR J1825$-$0935 also shows a correlation 
  between the pulse shape and spin-down rate. \citet{lhk+10} reported that pulse-shape changes
  between the Q-mode and the B-mode are highly correlated with the spin-down rate changes
  in PSR J1825$-$0935.

The most remarkable feature of PSR J1825$-$0935 is the peculiar and significant
anticorrelation between the intensity of the IP and the PC. Observations revealed
that the IP is visible when the PC is weak or absent (i.e., in the Q-mode);
on the other hand, the IP becomes very weak or undetectable when the PC is present
(i.e., in the B-mode) \citep{fw82,gjk+94}. Such an anticorrelation is naturally expected
in models where the PC and the IP are generated at the same magnetic
pole. \citet{dzg05} proposed that the IP and the PC originate from the same emission
region and that the PC emission intermittently reverses its direction to form the IP.
Although it is much harder to understand the IP-PC correlation in two-pole models,
on the basis of the separation between the IP and the MP and/or polarization properties
of this pulsar, many investigators argued that PSR J1825$-$0935 is an almost orthogonal
rotator and that the IP and the PC are emitted from two opposite magnetic poles
\citep{bmr10,mgr11,lmr12,hkh+17}.

Another curious property of PSR J1825$-$0935 is the periodic
modulation in the Q-mode.  Fluctuation-spectral analyses show that, in
the Q-mode, the IP and the MP are modulated at the same period and
their fluctuations are highly correlated
\citep{bmr10,lmr12}. \citet{lmr12} associated the periodic Q-mode
modulation with subpulse drifting. However, no organized drifting
pattern has been detected \citep{hkh+17}.

With high-sensitivity single-pulse observations, this paper aims to
explore previously unknown properties of the periodic Q-mode
modulation in PSR J1825$-$0935.  Details of the observations and data
processing are given in Section~\ref{sec:obs}. In
Section~\ref{sec:results} we present details of the observed periodic
intensity modulations.  The implications of the results are discussed
in Section~\ref{sec:discussion}.

\section{Observations} \label{sec:obs}

The observational data analyzed in this paper were downloaded from the
Parkes Pulsar Data Archive which is publicly available
online\footnote{\url{https://data.csiro.au}} \citep{hmm+11}.  The
single-pulse observations were made using the Parkes 64-m radio
telescope at six epochs with the center beam of the 20~cm Multibeam
receiver \citep{swb+96} and the Parkes digital filterbank systems
PDFB3 and PDFB4. See \citet{mhb+13} for further details of the
receiver and backend systems.  For the observations reported here, the
total bandwidth was 256 MHz centred at 1369 MHz with 512 channels
across the band.  The observing dates were 2013 March 25, June 23,
2014 February 20, April 19, July 28 and August 20. For each
observation, the duration was 8 min and the sampling interval was 256
$\mu$s.

The data were first reduced using the {\tt\string DSPSR} package
\citep{vb11} to de-disperse and produce single-pulse integrations
which were recorded using the {\tt\string PSRFITS} data format
\citep{hvm04} with 1024 phase bins per rotation period. The pulsar's
rotational ephemeris was taken from the ATNF Pulsar Catalogue
V1.59\footnote{\url{http://www.atnf.csiro.au/research/pulsar/psrcat/}}
\citep{mhth05}. Strong narrow-band radio-frequency interference (RFI)
in the archive files was removed in affected frequency
channels. Broad-band impulsive RFI was also removed in affected time
sub-integrations. Following \citet{mlc+01}, the single-pulse
integrations were then corrected for an effective high-pass filter in
the search-mode quantization algorithm.  The corrected single-pulse
integrations were processed using the {\tt\string PSRCHIVE} software
package \citep{vb11}.  The analysis of fluctuation spectra was carried
out with the {\tt\string PSRSALSA} package \citep{wel16} which is
freely available
online\footnote{\url{https://github.com/weltevrede/psrsalsa}}.
Following \citet{ymv+11}, the flux density, polarization and pcm
calibration were carried out for one observation for later
polarization analysis. The rotation measure (RM) value was obtained
from \citet{jk18}. Polarization parameters are in accordance with the
astronomical conventions described by \citet{vmjr10}.

\section{Results} \label{sec:results}

\begin{figure}
\centering
\includegraphics[angle=0,width=\columnwidth]{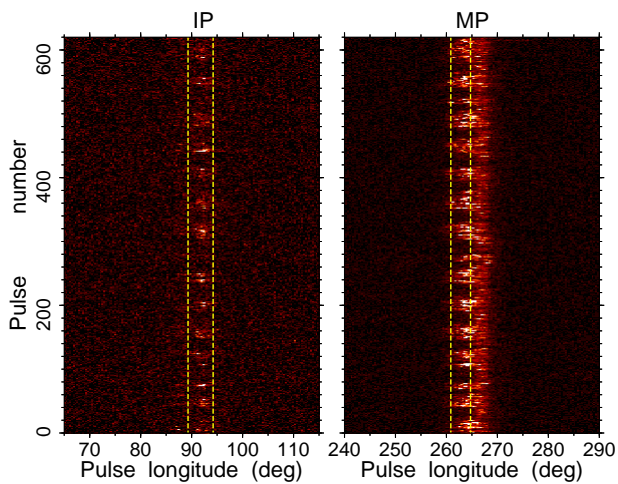}
\caption{A Q-mode single-pulse stack from the 2014 August 20 observation. The left and
         right panels show respectively the longitude range around
         the IP and the MP. The two vertical dashed lines in each panel define
         a pulse-phase window for computing the pulse energy used in
         Figure~\ref{fig:energy_var}. Note that, to see the IP more clearly,
         the range of the colour scale used in the left panel is
         different from that in the right panel.\label{fig:stack_q}}
\end{figure}

\begin{figure}
\centering
\includegraphics[angle=0,width=\columnwidth]{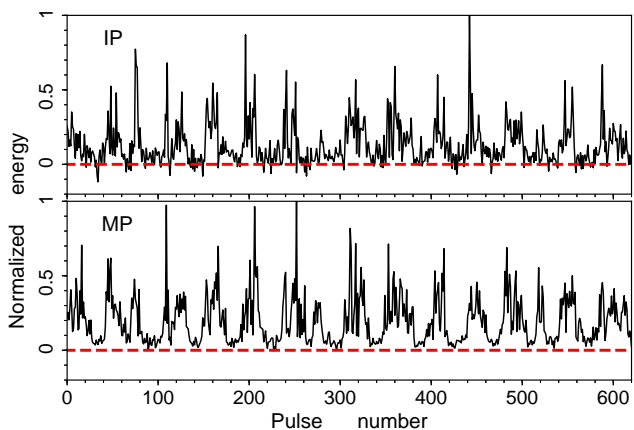}
\caption{Pulse energy variations for the IP (upper panel) and the leading component of the
         MP (lower panel) for the 2014 August 20 observation.
         The pulse phase windows for calculating the pulse energy of each
         component were defined by the vertical dashed lines in
         Figure~\ref{fig:stack_q}. \label{fig:energy_var}}
\end{figure}

The periodic fluctuation in the Q-mode of PSR J1825$-$0935, in which the
IP and the MP are modulated at the same period, is well known from
earlier studies.  To see this feature more clearly, we plotted the
pulse stack of the 2014 August 20 observation, which is a pure Q-mode
observation, in Figure~\ref{fig:stack_q}.  The left and right panels
show the single pulses of respectively the IP and MP for the same
rotation of the pulsar, where the IP is assumed to precede the MP.
The pulse energy variations with time of the IP and the leading region
of the MP for the same observation are presented in
Figure~\ref{fig:energy_var}. The pulse phase windows for calculating
the pulse energy were determined by eye (see
Figure~\ref{fig:stack_q}).  Figure~\ref{fig:stack_q} gives the
impression that the periodic modulation could be periodic nulling in
the IP and the leading component of the MP. However,
Figure~\ref{fig:energy_var} shows that the periodic modulation differs
from pulse nulling in two key aspects. Firstly, it is generally
believed that the transitions from bursts to nulls and the transitions
from nulls to bursts are abrupt (e.g., \citealt{wmj07}) rather than
gradual as in Figure~\ref{fig:energy_var}. Secondly, the pulse energy
effectively drops to zero in null states, but the pulse energy in
Figure~\ref{fig:energy_var} is often larger than zero even during the
apparent ``null'' states. We therefore suggest that the periodic
Q-mode modulation in PSR J1825$-$0935 is a periodic
longitude-stationary intensity modulation occurring in the interpulse
and the main pulse rather than periodic pulse nulling.  A similar
example is PSR B0826$-$34 whose weak mode had been regarded as a
complete null \citep{bmh+85}. However, a weak emission profile was
detected during the ``null'' states of the pulsar \citep{elg+05},
implying that the ``null'' state was a weak-emission mode rather than
being a null.

\subsection{Fluctuation spectra} \label{sec:fluc}

Figure~\ref{fig:stack_q} shows that the modulation period is about
40$P_1$, where $P_1$ is the basic pulse period, approximately
769~ms. \citet{bmr10} found the Q-mode modulation frequency to be
43$P_1$, while \citet{lmr12} gave a modulation frequency of
46.6$P_1$. To investigate the Q-mode modulation frequency further, we
carried out an analysis of fluctuation spectra by calculating the
longitude-resolved modulation index, the longitude-resolved
fluctuation spectrum \citep[LRFS,][]{bac70b} and the two-dimensional
fluctuation spectrum \citep[2DFS,][]{es02} for the observations
reported here. The longitude-resolved modulation index is a measure of
the amount of intensity variability at a given pulse longitude. The
LRFS is used to determine the presence of periodicities for each pulse
longitude bin.  The 2DFS is a different type of fluctuation spectrum
that can be used to identify if the emission drifts in pulse longitude
from pulse to pulse. Similar to the LRFS, the vertical frequency
axis of the 2DFS is expressed as $P_1/P_3$, where $P_3$ corresponds to
the $\sim$40$P_1$ periodicity of intensity modulation seen in the
single-pulse stack. The horizontal frequency axis of the 2DFS is
expressed as $P_1/P_2$, where $P_2$ corresponds to the characteristic
horizontal time separation between drifting bands.  For more details
about the analysis, see \citet{wes06}.

\begin{figure*}
\centering
\includegraphics[angle=0,width=0.95\textwidth]{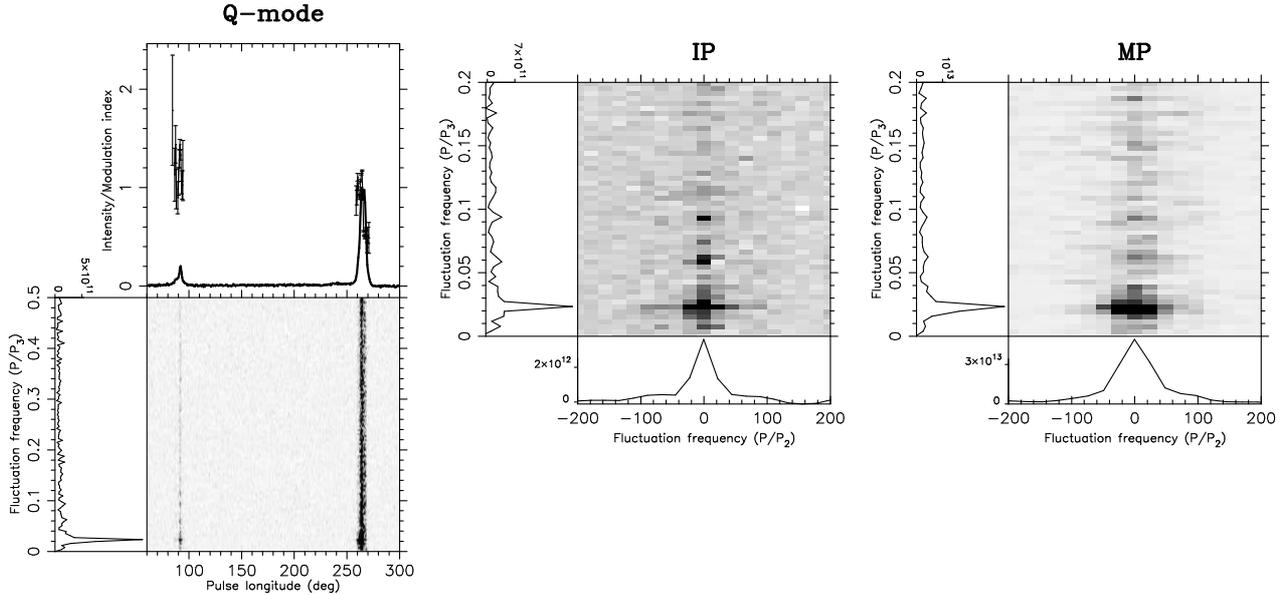}
\caption{Results of fluctuation analysis for the 2014 August 20 observation.
         The upper panel of the left column shows the mean pulse profile
         (solid line) and longitude-resolved   modulation index (points
         with error bars). The LRFS and a side panel showing the
         horizontally integrated power are given below this panel.
         The 2DFS and  side panels showing horizontally (left) and vertically
         (bottom) integrated power are plotted for the IP (middle column)
         and the MP (right column).
 \label{fig:fluc}}
\end{figure*}

Figure~\ref{fig:fluc} gives an example of our results based on the
observation of 2014 August 20. From Figure~\ref{fig:fluc}, we can see
that the modulation index of the leading component of the MP is higher
than that of the trailing component. This is not surprising given the
much stronger modulation visible in the leading component of the MP.
Examination of the LRFS (Figure~\ref{fig:fluc}) shows that, for both
the IP and the MP, the spectral peak occurs between 0.0234 and 0.0236
cycles per period (cpp), implying that both the IP and the MP show the
same $P_3$ of $42.6\pm 0.2\;P_1$, consistent with the results of
\citet{bmr10}. Fluctuation spectra of the six observations give
exactly the same value of $P_3$, and therefore we believe that the
modulation period for PSR J1825$-$0935 is quite stable on a timescale
of several years. However, the value of $P_3$ reported by
  \citet{lmr12} is somewhat larger than ours, this indicates that there
may exist fluctuations in the modulation period on a longer timescale.
Furthermore, the 2DFS is perfectly symmetric about
the vertical axis for each component. This indicates that subpulses in
successive pulses do not drift on average to later or earlier pulse
longitudes in either the IP or the MP.  There are therefore no drifting
subpulses in the Q-mode.

\subsection{Phase-locking} \label{sec:locking}

\begin{figure}
\centering
\includegraphics[angle=0,width=\columnwidth]{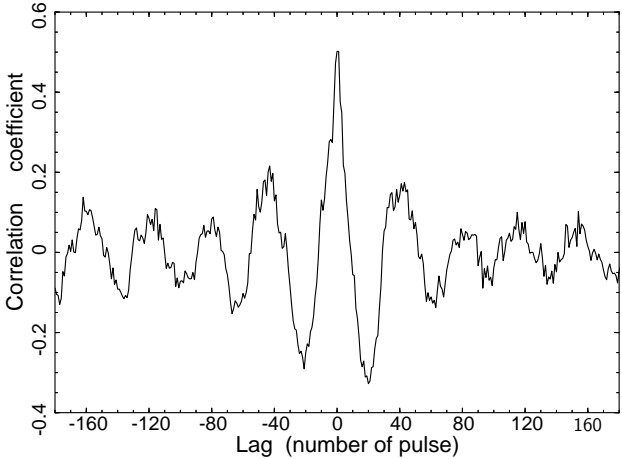}
\caption{The normalized cross-correlation between the pulse energies of the IP
  and that of the leading component of the following MP for the 2014 August 20
  observation. 
  \label{fig:energy_corr}}
\end{figure}

A phase-locked relationship between the IP and the MP modulation
patterns of PSR J1825$-$0935 had been found by earlier
investigators. \citet{bmr10} showed that there is a significant
correlation between pulse sequences of the 2$P_1$ delayed IP and the
leading component of the MP.  \citet{lmr12} argued that the IP and the
MP pulse sequences are offset in modulation phase by approximately 12
rotation periods. We investigated if there are any phase relations
between the 42.6$P_1$ modulation patterns observed in the IP and the
MP by correlating the measured pulse energies for successive pulses in
the IP and the MP of the observation of 2014 August 20. The cross-correlation
function (XCF), using the same pulse
longitude regions as in Figure~\ref{fig:energy_var}, is shown in
Figure~\ref{fig:energy_corr}. The XCF 
reaches a maximum value of 0.502 at zero lag, implying that a phase-locked 
relationship exists between the IP and the following MP. A T-test
shows that, at the zero lag, the correlation between the IP intensity and the
following MP intensity is significant at 0.01 level. The modulation periodicity
is seen in the XCF at delays of about $\pm 40 P_1$ and
multiples of this. Correlations are smaller than the zero-lag value
for all other lags. There is no sign of a significant XCF feature at
either $\pm 2P_1$ or $\pm 12P_1$. 

\subsection{State separation} \label{sec:mode}

\begin{figure*}
\centering
\includegraphics[angle=0,width=0.85\textwidth]{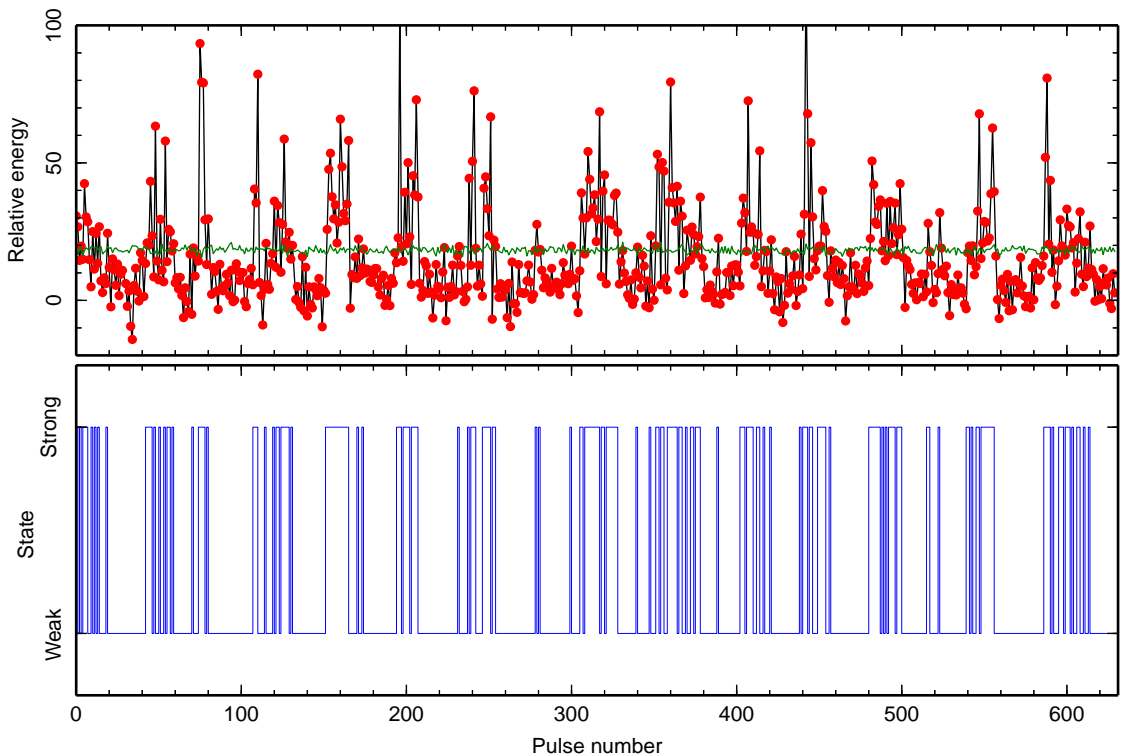}
\caption{The pulse energy sequence for the IP pulses (upper
         panel) with their corresponding identified strong/weak state
         (lower panel) for the 2014 August 20 observation.
         In the upper panel, a red dot is the on-pulse energy
         of a given IP pulse and the green line represents the three times level of
         the rms of the off-pulse region for single pulses. Pulses for which pulse
         energy below the green line were marked as weak-state pulses and pulses
        above the green line were marked as strong-state pulses.
\label{fig:mode}}
\end{figure*}

As a result of the periodic intensity modulation in the Q-mode, the IP
and the leading component of the MP switch between a strong state and a
weak state periodically (see Figure~\ref{fig:energy_var}).
In this paper, the term ``strong state'' refers to the strong state of the
Q-mode, and the term ``weak state'' refers to the weak state of the Q-mode.
Identifying the two states is required for further investigation. This is done by
comparing the on-pulse energy of the IP of individual pulses with the
system noise level. The uncertainty in the on-pulse energy of the IP
$\sigma_{\rm IP,on}$ is given by $\sqrt{N_{\rm on}}\sigma_{\rm off}$
\citep[cf.][]{bgg10}, where $N_{\rm on}$ is the number of on-pulse
longitude bins of the IP which was estimated from the mean pulse
profile and $\sigma_{\rm off}$ is the rms of the off-pulse region for
single pulses. We classified pulses with on-pulse energy of the IP
smaller than $3\sigma_{\rm IP,on}$ as weak-state pulses and the others as
strong-state pulses.
Figure~\ref{fig:mode} shows the results of the state separation
for the 2014 August 20 observation.
By performing the state separation in this way for six
Q-mode observations, we found that 69\% of the Q-mode time for PSR
J1825$-$0935 was in the weak state and 31\% was in the strong state.

\subsection{Polarization} \label{sec:poln}
To compare the weak state and the strong state further, we
analysed the polarization properties for both states. Of the six observations,
only the 2014 April 19 observation has suitable flux and polarization calibration
data. Most of the six observations show pure Q-mode emission,
however occasional short-duration B-mode emission occurs in the 2014 April 19
observation. To investigate the polarization properties of the weak and strong
states, B-mode emission must be removed.
As the most remarkable feature of B-mode is the presence of a strong PC component,
in order to remove the B-mode pulses from that observation, we
calculated the peak signal-to-noise ratio (S/N) for the PC region.
That is, the peak S/N is calculated as the ratio between the maximum intensity amplitude
of the PC on-pulse region, which was estimated from the B-mode mean pulse profile,
in a given rotation period and the standard deviation of the baseline points in the same
period \citep{ywm+18}. Pulses
with peak S/N larger than 3 were identified as B-mode pulses and then removed from
the data. The total number of the resulting Q-mode pulses for the 2014 April 19
observation is 538, which is about 86\% of the whole observation.

\begin{figure*}
\centering
\includegraphics[angle=0,width=0.95\textwidth]{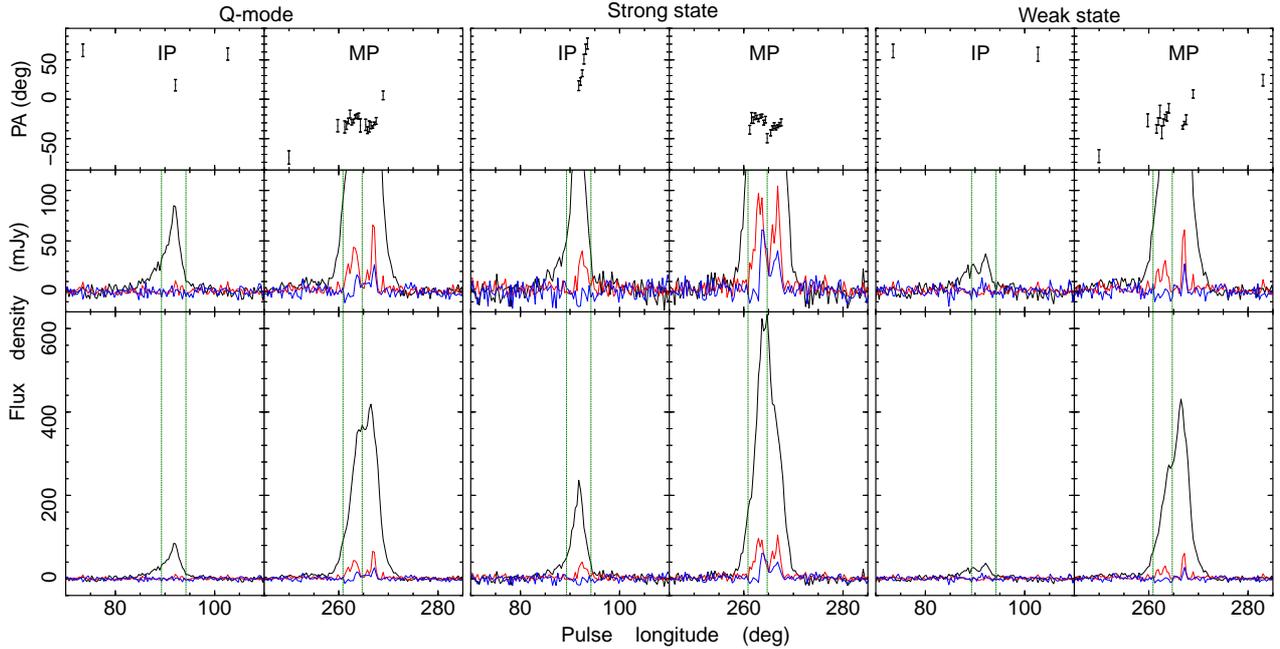}
\caption{Polarization profiles for
        the Q-mode pulses (left), the strong-state
        pulses (middle) and the weak-state pulses (right) for the
        interpulse and main pulse from the 2014 April
        19 observation. The bottom panels show the pulse profiles for total intensity
        (black line), linearly polarized intensity (red line), and circularly polarized
        intensity (blue line). The middle panels give expanded plots showing
        low-level details of the polarization profiles and the top panels give the
        position angles of the linearly polarized emission. The two vertical dashed lines
        in each panel define the same pulse-phase window as
        Figure~\ref{fig:stack_q}.
         \label{fig:poln}}
     \end{figure*}

We show polarization profiles for the 2014 April 19 observation in
Figure~\ref{fig:poln}, and Table~\ref{tab:poln} gives a summary of the
polarization parameters for different states. The mean flux density
$S$, the mean linear polarization intensity
$\left.\left\langle{L}\right\rangle\right.$, the mean circular
polarization intensity $\left.\left\langle{V}\right\rangle\right.$ and
the mean absolute circular polarization intensity
$\left.\left\langle{|V|}\right\rangle\right.$ were all averaged over
the IP/MP on-pulse window.  The pulse
energy of the IP in the strong state is four times larger than that
in the weak state. The MP in the weak state is also relatively weak.
The MP has a very strong leading component in the strong state, while
the leading component becomes much weaker than the trailing component
in the weak state.  The fractional linear polarization in the IP of
the strong state is higher than that of the weak state, while the
fractional circular polarization in the IP of the weak state is a
little higher. In the strong state, the circular polarization
intensity of the leading component of the MP is stronger than the
trailing component. But the circular polarization intensity of the
leading component is almost negligible in the weak state. In agreement
with \citet{jk18}, the strong state shows a rapid swing of
polarization position angle in its IP.

\begin{figure}
\centering
\includegraphics[angle=0,width=0.9\columnwidth]{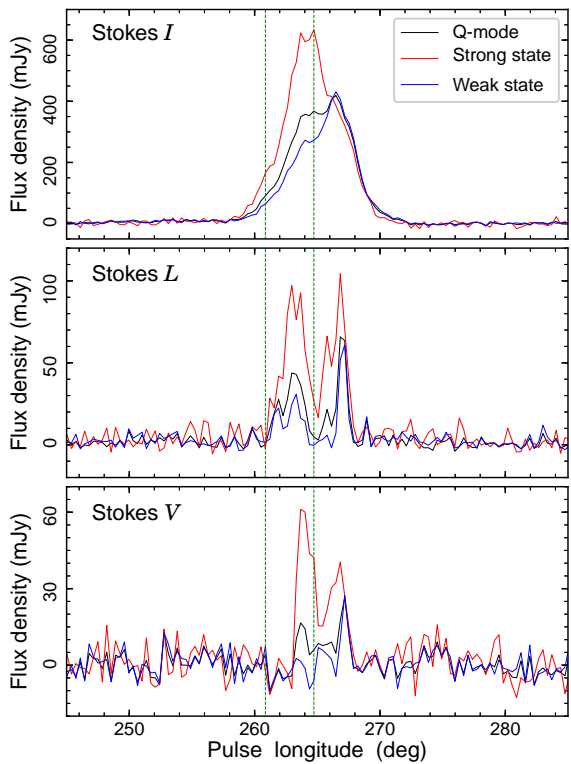}
\caption{The comparison of the polarization properties of the MP in
  the strong state and the weak state. The top, middle and bottom
  panels show respectively the total intensity, linearly polarized
  intensity and circularly polarized intensity for different
  states. The two vertical dashed lines in each panel define the same
  pulse-phase window as the right panel of Figure~\ref{fig:stack_q}.
         \label{fig:comp}}
     \end{figure}

Figure~\ref{fig:poln} also shows that, although the trailing edge of
the MP of the pulse profile is very similar in the weak and strong
states, the polarization of the trailing component of the MP is quite
different in the two states. To see the difference more clearly, we
compare total intensity, linearly polarized intensity and circularly
polarized intensity of the two states in Figure~\ref{fig:comp}. We can
clearly see that both the linear polarization and the circular
polarization are relatively stronger in the strong state across the
whole MP on-pulse area.  This suggests that the trailing component of
the MP must also be modulated, that is, the periodic modulation does
involve other parts of the magnetosphere, not just the parts where the
total intensity is modulated.

\begin{table*}
\begin{center}
\centering
\begin{minipage}[]{140mm}
\caption{Flux density and polarization parameters of the IP and the MP for the
strong and weak states for the 2014 April 19 observation.}
\label{tab:poln}
\end{minipage}
\begin{tabular}{cccccccccccc}
\hline
State  &\multicolumn{5}{c}{IP} &  & \multicolumn{5}{c}{MP}\\
\cline{2-6}
\cline{8-12}
  &  $S$ & $W_{50}$  &$\left.\left\langle{L}\right\rangle\middle/S\right.$
  &$\left.\left\langle{V}\right\rangle\middle/S\right.$
  &$\left.\left\langle\left|{V}\right|\right\rangle\middle/S\right.$  &  &$S$ & $W_{50}$  &$\left.\left\langle{L}\right\rangle\middle/S\right.$ &$\left.\left\langle{V}\right\rangle\middle/S\right.$ &$\left.\left\langle\left|{V}\right|\right\rangle\middle/S\right.$ \\
  & (mJy)   &($\degr$)  &(\%)  &(\%)  &(\%) &  & (mJy)   &($\degr$)  &(\%)  &(\%)  &(\%) \\
\hline
Q-mode       &29.4  &2.9  &2.3   &$-2.7$  &11.4  &  &163.9  &5.7  &7.7   &1.5  &3.3 \\
Strong  state  &67.2  &2.5  &12.6  &$-3.1$  &9.3  &  &218.6  &5.0  &12.8  &4.9  &6.1 \\
Weak  state    &15.8  &5.3  &4.1   &3.6     &23.8  &  &144.4  &4.9  &6.0   &0.2   &3.3 \\
\hline
\end{tabular}
\end{center}
\end{table*}

\section{Discussion and conclusions} \label{sec:discussion}

The periodic Q-mode modulation of PSR J1825$-$0935 was discovered more
than twenty years ago \citep{gjk+94}, but its nature remains
unclear. \citet{lmr12} related the intensity modulation to drifting
subpulses. However, no organized drifting pattern has been observed in
observational data.  Recently, several periodic non-drift amplitude
fluctuation phenomena have been observed in radio pulsar emission.
\citet{mr17} reported similar longitude-stationary
modulation in PSR B1946+35 and they compared modulation in this pulsar
with other pulsar modulation phenomena. It was shown that the periodic
non-drift amplitude modulation is very different from subpulse
drifting and thus was considered to be a new phenomenon.
\citet{bmm+16} detected periodic features in 57 pulsars including 29
pulsars that show no systematic drift but periodic amplitude
fluctuation. They found a relationship between
the drifting phenomenon and the spin-down luminosity $\dot{E}$.
In those 57 pulsars, pulsars showing subpulse drifting are seen to
lie below $\dot{E}\simeq 2\times$10$^{32}\ {\rm erg\,s^{-1}}$,
whereas other pulsars with $\dot{E}$ above this value
all showed non-drift amplitude modulation. PSR J1825$-$0935 has an
$\dot{E}\sim$ $4.6\times$10$^{33}\ {\rm erg\,s^{-1}}$ \citep{ywml10}, and
thus lies in the non-drift amplitude modulation group. 

Another similar phenomenon is periodic nulling that has been
widely reported in many pulsars
\citep{hr07,hr09,rw08,rwb13,gjw14,gyy+17,bmm17}. The periodicities of
the amplitude modulation are very similar to periodic nulling, causing
\citet{bmm17} to propose that they both may have a common origin.
\citet{hr07} used the rotating subbeam carousel model to account for
periodic nulling, in which the conal subbeams rotate around the
central core component. When the line of sight passes through the
empty region between two conal subbeams or through a extinguished
subbeam, nulling occurs, and this can repeat after certain spin
periods of the pulsar. However, \citet{bmm17} found that the missing
line of sight model did not adequately explain the periodic nulling
observed in many pulsars, particularly in pulsars with core
components.  Additionally, because of lack of phase memory and partial
nulls, the missing line of sight model is probably not applicable to
the quasi-periodic feature seen in PSR J1741$-$0840 \citep{gyy+17}.
For the periodic Q-mode modulation of 
PSR J1825$-$0935 reported in this paper, the IP and the MP switch between a
strong state and a weak state periodically. The weak state shows relatively
weak emission in the IP and the leading component of the MP, but not complete
nulls. This does not fit very well with the missing line of sight model.
Furthermore, as is shown in Figure~\ref{fig:comp}, the trailing edge of the
MP of the pulse profile is similar in the weak and strong states, but the
polarization of the trailing component of the MP differs significantly in the two
states. The polarization difference is not readily explained with the rotating
subbeam carousel model.

Our results show that the periodic Q-mode modulation of PSR
J1825$-$0935 is caused by regular switching between two emission
states. It is possible that the periodic modulation is actually a mode
change. However, in most pulsars, mode changing is not periodic and so
the relationship between these two phenomena is somewhat
unclear. Never-the-less, we argue that the periodic modulations in the
Q-mode of PSR J1825$-$0935 result from similarly periodic fluctuations
in the magnetospheric field currents that are not directly related to
sub-pulse drifting.

In agreement with earlier results, we show that there is a
phase-locked relationship between the IP and MP modulations in the
Q-mode of PSR J1825$-$0935. Similar phase-locked features were also
detected in PSRs B1702$-$19 \citep{wws07} and B1055$-$52
\citep{wwj12}. No phase lag in the modulation patterns was found in
our data, implying that the IP and the following MP have correlated
intensity variations. The zero-lag phase-locked feature is found to be identical
in all six observations we used. This implies the phase-locked feature
is sustained over several years and probably a permanent feature of
PSR J1825$-$0935.

Recently, a close correlation between the emission variability and the
  rotational properties has been reported in a number of pulsars
  \citep{klo+06,lhk+10,ksj13,bkb+14,bkj+16,kyw+18}. Theses pulsars provide
  direct observational evidence for the connection between magnetosphere behaviour
  and rotational behaviour. For PSR J1825$-$0935, the variation of the ratio of the 
  amplitudes of the PC and MP (which indicates the Q-mode or B-mode) has been found to
  correlate with the variation of spin-down rate in a long timescale ($\sim$10 yr)
  \citep{lhk+10}. It is possible to determine the spin-down rates in the weak
  and strong states of the Q-mode separately by using standard pulsar timing techniques
  on each state. However, more single-pulse pure Q-mode 
  observational data spanning a long timescale are required for this study.

PSR J1825$-$0935 is commonly considered to be an orthogonal rotator
where the IP and MP are emitted from two opposite magnetic poles. In a
recent work, \citet{jk18} presented a high quality polarization
profile for PSR J1825$-$0935. A rapid swing of polarization position
angle is seen in both the main pulse and the interpulse providing new
evidence for the two-pole model. In the two-pole geometry, the same
modulation in the IP and MP of PSR J1825$-$0935 is puzzling, because
it requires information transfer between the two opposite magnetic
poles. Some investigators have suggested the possibility of global
change in the pulsar magnetosphere on very short timescales
\citep{lmr12}. Another possible explanation is that the two magnetic poles
are affected simultaneously through non-radial oscillations of the
body of the neutron star itself \citep{wws07}. At present, neither of
these explanations is very compelling and maybe the one-pole model of 
\citet{dzg05} should be further considered.  It is clear that more
investigation of possible emission models is needed to clarify the
mechanism(s) responsible for the observed phenomena.

\section*{Acknowledgements}

This work is supported by National Basic Research Program of China
(973 Program 2015CB857100), National Natural Science Foundation of China
(Nos. U1831102, U1731238, U1631106, 11873080, U1838109),
the Strategic Priority Research Program of Chinese Academy of Sciences
(No. XDB23010200), the National Key Research and Development Program of China
(No. 2016YFA0400800), the 2016 Project of Xinjiang Uygur Autonomous Region of
China for Flexibly Fetching in Upscale Talents and the
CAS “Light of West China” Program (2016-QNXZ-B-24 and 2017-XBQNXZ-B-022).
The Parkes radio telescope is
part of the Australia Telescope, which is funded by the Commonwealth of
Australia for operation as a National Facility managed by the Commonwealth
Scientific and Industrial Research Organisation.





\begin{thebibliography}{99}
\makeatletter
\relax
\def\mn@urlcharsother{\let\do\@makeother \do\$\do\&\do\#\do\^\do\_\do\%\do\~}
\def\mn@doi{\begingroup\mn@urlcharsother \@ifnextchar [ {\mn@doi@}
  {\mn@doi@[]}}
\def\mn@doi@[#1]#2{\def\@tempa{#1}\ifx\@tempa\@empty \href
  {http://dx.doi.org/#2} {doi:#2}\else \href {http://dx.doi.org/#2} {#1}\fi
  \endgroup}
\def\mn@eprint#1#2{\mn@eprint@#1:#2::\@nil}
\def\mn@eprint@arXiv#1{\href {http://arxiv.org/abs/#1} {{\tt arXiv:#1}}}
\def\mn@eprint@dblp#1{\href {http://dblp.uni-trier.de/rec/bibtex/#1.xml}
  {dblp:#1}}
\def\mn@eprint@#1:#2:#3:#4\@nil{\def\@tempa {#1}\def\@tempb {#2}\def\@tempc
  {#3}\ifx \@tempc \@empty \let \@tempc \@tempb \let \@tempb \@tempa \fi \ifx
  \@tempb \@empty \def\@tempb {arXiv}\fi \@ifundefined
  {mn@eprint@\@tempb}{\@tempb:\@tempc}{\expandafter \expandafter \csname
  mn@eprint@\@tempb\endcsname \expandafter{\@tempc}}}

\bibitem[\protect\citeauthoryear{Backer}{Backer}{1970}]{bac70b}
Backer D.~C.,  1970, Nature, 227, 692

\bibitem[\protect\citeauthoryear{{Backus}, {Mitra}  \& {Rankin}}{{Backus}
  et~al.}{2010}]{bmr10}
{Backus} I.,  {Mitra} D.,   {Rankin} J.~M.,  2010, \mn@doi [MNRAS]
  {10.1111/j.1365-2966.2009.16102.x}, \href
  {http://adsabs.harvard.edu/abs/2010MNRAS.404...30B} {404, 30}

\bibitem[\protect\citeauthoryear{{Basu}, {Mitra}, {Melikidze}, {Maciesiak},
  {Skrzypczak}  \& {Szary}}{{Basu} et~al.}{2016}]{bmm+16}
{Basu} R.,  {Mitra} D.,  {Melikidze} G.~I.,  {Maciesiak} K.,  {Skrzypczak} A.,
   {Szary} A.,  2016, \mn@doi [ApJ] {10.3847/1538-4357/833/1/29}, \href
  {http://adsabs.harvard.edu/abs/2016ApJ...833...29B} {833, 29}

\bibitem[\protect\citeauthoryear{{Basu}, {Mitra}  \& {Melikidze}}{{Basu}
  et~al.}{2017}]{bmm17}
{Basu} R.,  {Mitra} D.,   {Melikidze} G.~I.,  2017, \mn@doi [ApJ]
  {10.3847/1538-4357/aa862d}, \href
  {http://adsabs.harvard.edu/abs/2017ApJ...846..109B} {846, 109}

\bibitem[\protect\citeauthoryear{{Bhattacharyya}, {Gupta}  \&
  {Gil}}{{Bhattacharyya} et~al.}{2010}]{bgg10}
{Bhattacharyya} B.,  {Gupta} Y.,   {Gil} J.,  2010, \mn@doi [MNRAS]
  {10.1111/j.1365-2966.2010.17116.x}, \href
  {http://adsabs.harvard.edu/abs/2010MNRAS.408..407B} {408, 407}

\bibitem[\protect\citeauthoryear{Biggs, McCulloch, Hamilton, Manchester  \&
  Lyne}{Biggs et~al.}{1985}]{bmh+85}
Biggs J.~D.,  McCulloch P.~M.,  Hamilton P.~A.,  Manchester R.~N.,   Lyne
  A.~G.,  1985, MNRAS, 215, 281

\bibitem[\protect\citeauthoryear{{Brook}, {Karastergiou}, {Buchner}, {Roberts},
  {Keith}, {Johnston}  \& {Shannon}}{{Brook} et~al.}{2014}]{bkb+14}
{Brook} P.~R.,  {Karastergiou} A.,  {Buchner} S.,  {Roberts} S.~J.,  {Keith}
  M.~J.,  {Johnston} S.,   {Shannon} R.~M.,  2014, \mn@doi [ApJ]
  {10.1088/2041-8205/780/2/L31}, \href
  {http://adsabs.harvard.edu/abs/2014ApJ...780L..31B} {780, L31}

\bibitem[\protect\citeauthoryear{{Brook}, {Karastergiou}, {Johnston}, {Kerr},
  {Shannon}  \& {Roberts}}{{Brook} et~al.}{2016}]{bkj+16}
{Brook} P.~R.,  {Karastergiou} A.,  {Johnston} S.,  {Kerr} M.,  {Shannon}
  R.~M.,   {Roberts} S.~J.,  2016, \mn@doi [MNRAS] {10.1093/mnras/stv2715},
  \href {http://adsabs.harvard.edu/abs/2016MNRAS.456.1374B} {456, 1374}

\bibitem[\protect\citeauthoryear{Dyks, Zhang  \& Gil}{Dyks
  et~al.}{2005}]{dzg05}
Dyks J.,  Zhang B.,   Gil J.,  2005, ApJ, 626, L45

\bibitem[\protect\citeauthoryear{{Edwards} \& {Stappers}}{{Edwards} \&
  {Stappers}}{2002}]{es02}
{Edwards} R.~T.,  {Stappers} B.~W.,  2002, A\&A, 393, 733

\bibitem[\protect\citeauthoryear{{Esamdin}, {Lyne}, {Graham-Smith}, {Kramer},
  {Manchester}  \& {Wu}}{{Esamdin} et~al.}{2005}]{elg+05}
{Esamdin} A.,  {Lyne} A.~G.,  {Graham-Smith} F.,  {Kramer} M.,  {Manchester}
  R.~N.,   {Wu} X.,  2005, MNRAS, 356, 59

\bibitem[\protect\citeauthoryear{Fowler \& Wright}{Fowler \&
  Wright}{1982}]{fw82}
Fowler L.~A.,  Wright G. A.~E.,  1982, A\&A, 109, 279

\bibitem[\protect\citeauthoryear{Fowler, Wright  \& Morris}{Fowler
  et~al.}{1981}]{fwm81}
Fowler L.~A.,  Wright G. A.~E.,   Morris D.,  1981, A\&A, 93, 54

\bibitem[\protect\citeauthoryear{{Gajjar}, {Joshi}  \& {Wright}}{{Gajjar}
  et~al.}{2014}]{gjw14}
{Gajjar} V.,  {Joshi} B.~C.,   {Wright} G.,  2014, \mn@doi [MNRAS]
  {10.1093/mnras/stt2389}, \href
  {http://adsabs.harvard.edu/abs/2014MNRAS.439..221G} {439, 221}

\bibitem[\protect\citeauthoryear{{Gajjar}, {Yuan}, {Yuen}, {Wen}, {Liu}  \&
  {Wang}}{{Gajjar} et~al.}{2017}]{gyy+17}
{Gajjar} V.,  {Yuan} J.~P.,  {Yuen} R.,  {Wen} Z.~G.,  {Liu} Z.~Y.,   {Wang}
  N.,  2017, \mn@doi [ApJ] {10.3847/1538-4357/aa96ac}, \href
  {http://adsabs.harvard.edu/abs/2017ApJ...850..173G} {850, 173}

\bibitem[\protect\citeauthoryear{Gil et~al.,}{Gil et~al.}{1994}]{gjk+94}
Gil J.~A.,  et~al., 1994, A\&A, 282, 45

\bibitem[\protect\citeauthoryear{{Herfindal} \& {Rankin}}{{Herfindal} \&
  {Rankin}}{2007}]{hr07}
{Herfindal} J.~L.,  {Rankin} J.~M.,  2007, \mn@doi [MNRAS]
  {10.1111/j.1365-2966.2007.12089.x}, \href
  {http://adsabs.harvard.edu/abs/2007MNRAS.380..430H} {380, 430}

\bibitem[\protect\citeauthoryear{{Herfindal} \& {Rankin}}{{Herfindal} \&
  {Rankin}}{2009}]{hr09}
{Herfindal} J.~L.,  {Rankin} J.~M.,  2009, \mn@doi [MNRAS]
  {10.1111/j.1365-2966.2008.14119.x}, \href
  {http://adsabs.harvard.edu/abs/2009MNRAS.393.1391H} {393, 1391}

\bibitem[\protect\citeauthoryear{{Hermsen} et~al.,}{{Hermsen}
  et~al.}{2017}]{hkh+17}
{Hermsen} W.,  et~al., 2017, \mn@doi [MNRAS] {10.1093/mnras/stw3135}, \href
  {http://adsabs.harvard.edu/abs/2017MNRAS.466.1688H} {466, 1688}

\bibitem[\protect\citeauthoryear{{Hobbs} et~al.,}{{Hobbs}
  et~al.}{2011}]{hmm+11}
{Hobbs} G.,  et~al., 2011, \mn@doi [PASA] {10.1071/AS11016}, \href
  {http://adsabs.harvard.edu/abs/2011PASA...28..202H} {28, 202}

\bibitem[\protect\citeauthoryear{{Hotan}, {van Straten}  \&
  {Manchester}}{{Hotan} et~al.}{2004}]{hvm04}
{Hotan} A.~W.,  {van Straten} W.,   {Manchester} R.~N.,  2004, PASA, 21, 302

\bibitem[\protect\citeauthoryear{{Johnston} \& {Kerr}}{{Johnston} \&
  {Kerr}}{2018}]{jk18}
{Johnston} S.,  {Kerr} M.,  2018, \mn@doi [MNRAS] {10.1093/mnras/stx3095},
  \href {http://adsabs.harvard.edu/abs/2018MNRAS.474.4629J} {474, 4629}

\bibitem[\protect\citeauthoryear{{Keith}, {Shannon}  \& {Johnston}}{{Keith}
  et~al.}{2013}]{ksj13}
{Keith} M.~J.,  {Shannon} R.~M.,   {Johnston} S.,  2013, \mn@doi [MNRAS]
  {10.1093/mnras/stt660}, \href
  {http://adsabs.harvard.edu/abs/2013MNRAS.432.3080K} {432, 3080}

\bibitem[\protect\citeauthoryear{{Kou}, {Yuan}, {Wang}, {Yan}  \& {Dang}}{{Kou}
  et~al.}{2018}]{kyw+18}
{Kou} F.~F.,  {Yuan} J.~P.,  {Wang} N.,  {Yan} W.~M.,   {Dang} S.~J.,  2018,
  \mn@doi [MNRAS] {10.1093/mnrasl/sly068}, \href
  {http://adsabs.harvard.edu/abs/2018MNRAS.478L..24K} {478, L24}

\bibitem[\protect\citeauthoryear{{Kramer}, {Lyne}, {O'Brien}, {Jordan}  \&
  {Lorimer}}{{Kramer} et~al.}{2006}]{klo+06}
{Kramer} M.,  {Lyne} A.~G.,  {O'Brien} J.~T.,  {Jordan} C.~A.,   {Lorimer}
  D.~R.,  2006, Science, 312, 549

\bibitem[\protect\citeauthoryear{{Latham}, {Mitra}  \& {Rankin}}{{Latham}
  et~al.}{2012}]{lmr12}
{Latham} C.,  {Mitra} D.,   {Rankin} J.,  2012, \mn@doi [MNRAS]
  {10.1111/j.1365-2966.2012.21985.x}, \href
  {http://adsabs.harvard.edu/abs/2012MNRAS.427..180L} {427, 180}

\bibitem[\protect\citeauthoryear{{Lyne}, {Hobbs}, {Kramer}, {Stairs}  \&
  {Stappers}}{{Lyne} et~al.}{2010}]{lhk+10}
{Lyne} A.,  {Hobbs} G.,  {Kramer} M.,  {Stairs} I.,   {Stappers} B.,  2010,
  \mn@doi [Science] {10.1126/science.1186683}, \href
  {http://adsabs.harvard.edu/abs/2010Sci...329..408L} {329, 408}

\bibitem[\protect\citeauthoryear{{Maciesiak}, {Gil}  \& {Ribeiro}}{{Maciesiak}
  et~al.}{2011}]{mgr11}
{Maciesiak} K.,  {Gil} J.,   {Ribeiro} V.~A.~R.~M.,  2011, \mn@doi [MNRAS]
  {10.1111/j.1365-2966.2011.18471.x}, \href
  {http://adsabs.harvard.edu/abs/2011MNRAS.414.1314M} {414, 1314}

\bibitem[\protect\citeauthoryear{Manchester et~al.,}{Manchester
  et~al.}{2001}]{mlc+01}
Manchester R.~N.,  et~al., 2001, MNRAS, 328, 17

\bibitem[\protect\citeauthoryear{{Manchester}, {Hobbs}, {Teoh}  \&
  {Hobbs}}{{Manchester} et~al.}{2005}]{mhth05}
{Manchester} R.~N.,  {Hobbs} G.~B.,  {Teoh} A.,   {Hobbs} M.,  2005, AJ, 129,
  1993

\bibitem[\protect\citeauthoryear{{Manchester} et~al.,}{{Manchester}
  et~al.}{2013}]{mhb+13}
{Manchester} R.~N.,  et~al., 2013, \mn@doi [PASA] {10.1017/pasa.2012.017},
  \href {http://adsabs.harvard.edu/abs/2013PASA...30...17M} {30, e017}

\bibitem[\protect\citeauthoryear{{Mitra} \& {Rankin}}{{Mitra} \&
  {Rankin}}{2017}]{mr17}
{Mitra} D.,  {Rankin} J.,  2017, \mn@doi [MNRAS] {10.1093/mnras/stx814}, \href
  {http://adsabs.harvard.edu/abs/2017MNRAS.468.4601M} {468, 4601}

\bibitem[\protect\citeauthoryear{Morris, Graham  \& Bartel}{Morris
  et~al.}{1981}]{mgb81}
Morris D.,  Graham D.~A.,   Bartel N.,  1981, MNRAS, 194, 7P

\bibitem[\protect\citeauthoryear{{Rankin} \& {Wright}}{{Rankin} \&
  {Wright}}{2008}]{rw08}
{Rankin} J.~M.,  {Wright} G.~A.~E.,  2008, \mn@doi [MNRAS]
  {10.1111/j.1365-2966.2008.13001.x}, \href
  {http://adsabs.harvard.edu/abs/2008MNRAS.385.1923R} {385, 1923}

\bibitem[\protect\citeauthoryear{{Rankin}, {Wright}  \& {Brown}}{{Rankin}
  et~al.}{2013}]{rwb13}
{Rankin} J.~M.,  {Wright} G.~A.~E.,   {Brown} A.~M.,  2013, \mn@doi [MNRAS]
  {10.1093/mnras/stt739}, \href
  {http://adsabs.harvard.edu/abs/2013MNRAS.433..445R} {433, 445}

\bibitem[\protect\citeauthoryear{Staveley-Smith et~al.,}{Staveley-Smith
  et~al.}{1996}]{swb+96}
Staveley-Smith L.,  et~al., 1996, PASA, 13, 243

\bibitem[\protect\citeauthoryear{{van Straten} \& {Bailes}}{{van Straten} \&
  {Bailes}}{2011}]{vb11}
{van Straten} W.,  {Bailes} M.,  2011, \mn@doi [PASA] {10.1071/AS10021}, \href
  {http://adsabs.harvard.edu/abs/2011PASA...28....1V} {28, 1}

\bibitem[\protect\citeauthoryear{{van Straten}, {Manchester}, {Johnston}  \&
  {Reynolds}}{{van Straten} et~al.}{2010}]{vmjr10}
{van Straten} W.,  {Manchester} R.~N.,  {Johnston} S.,   {Reynolds} J.~E.,
  2010, \mn@doi [PASA] {10.1071/AS09084}, \href
  {http://adsabs.harvard.edu/abs/2010PASA...27..104V} {27, 104}

\bibitem[\protect\citeauthoryear{{Wang}, {Manchester}  \& {Johnston}}{{Wang}
  et~al.}{2007}]{wmj07}
{Wang} N.,  {Manchester} R.~N.,   {Johnston} S.,  2007, \mn@doi [MNRAS]
  {10.1111/j.1365-2966.2007.11703.x}, 377, 1383

\bibitem[\protect\citeauthoryear{{Weltevrede}}{{Weltevrede}}{2016}]{wel16}
{Weltevrede} P.,  2016, \mn@doi [A\&A] {10.1051/0004-6361/201527950}, \href
  {http://adsabs.harvard.edu/abs/2016A%26A...590A.109W} {590, A109}

\bibitem[\protect\citeauthoryear{{Weltevrede}, {Edwards}  \&
  {Stappers}}{{Weltevrede} et~al.}{2006}]{wes06}
{Weltevrede} P.,  {Edwards} R.~T.,   {Stappers} B.~W.,  2006, A\&A, 445, 243

\bibitem[\protect\citeauthoryear{Weltevrede, Wright  \& Stappers}{Weltevrede
  et~al.}{2007}]{wws07}
Weltevrede P.,  Wright G.~A.~E.,   Stappers B.~W.,  2007, A\&A, 467, 1163

\bibitem[\protect\citeauthoryear{{Weltevrede}, {Wright}  \&
  {Johnston}}{{Weltevrede} et~al.}{2012}]{wwj12}
{Weltevrede} P.,  {Wright} G.,   {Johnston} S.,  2012, \mn@doi [MNRAS]
  {10.1111/j.1365-2966.2012.21207.x}, \href
  {http://adsabs.harvard.edu/abs/2012MNRAS.424..843W} {424, 843}

\bibitem[\protect\citeauthoryear{{Yan} et~al.,}{{Yan} et~al.}{2011}]{ymv+11}
{Yan} W.~M.,  et~al., 2011, \mn@doi [MNRAS] {10.1111/j.1365-2966.2011.18522.x},
  \href {http://adsabs.harvard.edu/abs/2011MNRAS.414.2087Y} {414, 2087}

\bibitem[\protect\citeauthoryear{{Yan}, {Wang}, {Manchester}, {Wen}  \&
  {Yuan}}{{Yan} et~al.}{2018}]{ywm+18}
{Yan} W.~M.,  {Wang} N.,  {Manchester} R.~N.,  {Wen} Z.~G.,   {Yuan} J.~P.,
  2018, \mn@doi [MNRAS] {10.1093/mnras/sty470}, \href
  {http://adsabs.harvard.edu/abs/2018MNRAS.476.3677Y} {476, 3677}

\bibitem[\protect\citeauthoryear{{Yuan}, {Wang}, {Manchester}  \& {Liu}}{{Yuan}
  et~al.}{2010}]{ywml10}
{Yuan} J.~P.,  {Wang} N.,  {Manchester} R.~N.,   {Liu} Z.~Y.,  2010, \mn@doi
  [MNRAS] {10.1111/j.1365-2966.2010.16272.x}, \href
  {http://adsabs.harvard.edu/abs/2010MNRAS.404..289Y} {404, 289}

\makeatother
\end{thebibliography}



\bsp	
\label{lastpage}
\end{document}